\documentclass[nofootinbib,letterpaper,aps,amsmath,amsfonts,twocolumn]{revtex4}
\usepackage[mathscr]{eucal}
\usepackage{graphicx}
\usepackage{amssymb}
\newcommand{\td}{\textup{d}}  
\newcommand{\del}{\partial}   
\def\eqref#1{(\ref{#1})}      
\def\er#1{eqn.\eqref{#1}}     
\def\nn{\nonumber}
\begin{document}
\title{All Conformal Effective String Theories are Isospectral to Nambu-Goto Theory}
\author{N.D. Hari Dass}
\affiliation{Centre for High-Energy Physics, Indian Institute of Science, Bangalore, India }
\email{dass@cts.iisc.ernet.in}
\affiliation{Poornaprajna Institute of Scientific Research, Bangalore}
\begin{abstract}
It is shown that all Polchinski-Strominger effective string theories are \emph{isospectral} to Nambu-Goto theory.
The relevance of these results to QCD-Strings is discussed.

\end{abstract}

\maketitle


\section{Introduction}
\label{intro}
String-like defects or solitons occur in a wide variety of physical systems. Some well-known examples are vortices
in superfluids, the Nielsen-Olesen vortices of quantum field theories, vortices in Bose-Einstein condensates and
QCD-strings(for reviews see \cite{balikuti}). Under suitable conditions these objects can behave quantum-mechanically. The
challenge then is to find consistent quantum mechanical descriptions in \emph{arbitrary}
dimensions. It should be recalled that fundamental string theories are consistent only in the so called critical
dimensions. Polyakov \cite{polya} gave formulations of string theories in \emph{sub-critical} dimensions; 
his ideas play a central role in what follows.
A pragmatic approach to such a quantum description would be to treat these objects  in some effective manner much as interactions
of pions are so succesfully described in terms of chiral effective field theories without any pretenses about them 
being fundamental at all scales.

Two such approaches to effective string theories exist in the literature. One due to L\"uscher and collaborators \cite{lwearly}, is 
entirely in terms of the $D-2$ \emph{transverse} physical degrees of freedom. It is a case where the \emph{gauge} is fixed completely
without any \emph{residual} invariance left. 
The other approach is the one pioneered by Polchinski and Strominger \cite{ps} where the theories are invariant under 
\emph{conformal
transformations} and the physical states are obtained by requiring that the generators of conformal transformations
annihilate them. These too are gauge-fixed theories but with leftover residual invariances characterized by conformal
transformations. The physical basis of both approaches is that the physical degrees of freedom are 
transverse. In the light of the results obtained in this paper, it has become important to reexamine this physical basis.

L\"uscher was the first to show \cite{lwearly} that the leading correction to ground state energy of a (closed)string
of length $2\pi R$ takes the form $V(R) = \sigma R -\frac{(D-2)\pi}{24}$. He drew attention to the fact that the leading correction, subsequently named the L\"uscher term,
is \emph{universal} depending only on space-time dimensionality. Polchinski and Strominger \cite{ps} showed, through explicit
construction of an approximately conformally invariant action for effective string theories that not only can string-like defects
be quantized in arbitrary dimensions but also that the leading correction is the L\"uscher term. Very recently 
L\"uscher and Weisz, using their path-breaking multilevel algorithm, 
 showed clear numerical evidence for this term \cite{lwrecent} in Lattice QCD. Subsequent large scale numerical 
simulations of the string-like behaviour of Yang-Mills flux tubes by Hari Dass and Majumdar \cite{ournum} pointed to the strong possibility that even the subleading $R^{-3}$ terms in  $D=3$ and $D=4$ were
universal and what is more, coincided with similar terms of Nambu-Goto theory \cite{Arv}. 

This rather unexpected result was analytically explained by Drummond and, by Hari Dass and Matlock\cite{drumorig,ouruniv}. 
Focus then 
shifted to finding ways of understanding even higher order corrections, and possibly an analysis to all orders. The first result we
obtained in this direction was the proof that  the action that Polchinski and Strominger used in \cite{ps} extended
to be \emph{exactly} conformally invariant \emph{to all orders} in \cite{covariant}, has the same spectrum as the Nambu-Goto theory
\cite{liou-all} to all orders. We had called this extended action by the name \emph{Polyakov-Liouville} action. Drummond \cite{drumorig} had shown that the next level at which candidate actions could be found
was only at $R^{-6}$.However, he did not identify the conformal transformations leaving invariant his actions, four in number, which we
have called \emph{Drummond Actions}. With the help of our covariant formalism we had shown that only two of these are linearly
independent and we had identified their invariance transformation laws \cite{covariant}. In the next all order result we had
shown that effective string actions with these Drummond terms also do not change the spectrum! \cite{drumacts}. In this paper we extend our proof 
to \emph{all classes} of Polchinski-Strominger effective string theories to all orders.

\section{Actions}
The total action for effective string theories has the form
\begin{equation}
\label{total}
S = S_0+S_\beta+\sum_j S_{cov}^j
\end{equation}
where
\begin{equation}
4\pi a^2~S_0 =  \int \td\tau^+ \td\tau^- L
 =  \int \td^2\tau  
  \del_+ {X^\mu} \del_- {X_\mu} 
\end{equation}
is the action for the free bosonic string theory and is consistent quantum mechanically only in $D=26$ space-time dimensions.
$S_\beta$ is the \emph{Polyakov-Liouville} action
\begin{equation}
\label{action2}
S_{\beta} = \frac{\beta}{4\pi} \int \td^2\tau \bigg\{
\frac {\del_+(\del_+X\cdot\del_-X)\del_-(\del_+X\cdot\del_-X)}{(\del_+X\cdot\del_-X)^2}
\bigg\}
.\end{equation}
$S_{\beta}$ is \emph{exactly}, in the sense to all orders in $R^{-1}$, invariant under
the conformal transformations
\begin{equation}
\label{urtrans}
\delta^{0}_\pm X^\mu = \epsilon^\pm(\tau^\pm)\del_\pm X^\mu
.\end{equation}
under which the leading action $S_0$ is also exactly invariant. The action and transformations laws originally used in 
\cite{ps} are related by a \emph{field redefinition} \cite{field}. The PS-action, equivalently $S_\beta$, is what yields
quantum consistency in \emph{all} dimensions. The so called  \emph{manifestly covariant}
actions $S_{cov}^j$ are such that their integrands transform as \emph{scalar densities}. They can be constructed systematically
with the help of the \emph{Covariant Calculus} developed in \cite{covariant}, and are therefore exactly invariant under
\er{urtrans}. An important difference from the algorithm given in \cite{ps} and elaborated in \cite{drumorig,ouruniv}
is that terms proportional to EOM can no longer be dropped as that would entail field redefinitions which could change
\er{urtrans}. In the next section we describe some essentials.
\section{Covariant Calculus}
\label{covMet}
\subsection{General Considerations}
\label{conf1}
We start with the general form of manifestly covariant action terms, more
specifically, terms that transform as scalar densities. A systematic
procedure for construction of such terms to any desired order in $1/R$
is given in \cite{covariant}.
\begin{equation}
I_{\textup{cov}} = \sqrt{g}D_{\alpha_1\beta_1..}X^{\mu_1}D_{\alpha_2\beta_2..}X^{\mu_2}\cdot A^{\alpha_1\beta_1\cdots\alpha_2\beta_2\cdots}B_{\mu_1\mu_2\cdots}
\end{equation}
where $A^{\alpha_1\beta_1\cdots\alpha_2\beta_2\cdots}$ is composed of
suitable factors of Levi-Civita and metric tensors on the
two-dimensional world sheet and $B_{\mu_1\mu_2\cdot}$ made up of
$\eta_{\mu\nu}$ and Levi-Civita tensors in target space. In the spirit of
the PS-construction, the covariant calculus is constructed based on the \emph{induced metric}
on the world-sheet given by $g_{\alpha\beta} = \del_\alpha X\cdot \del_\beta X$.
In the
conformal gauge, $g_{++} = g_{--} =0$,
this construction amounts to
stringing together a number of covariant derivatives so that there are
equal net numbers of $(+,-)$ indices, and use sufficient
inverse powers of $g_{+-}\equiv L = \del_+X\cdot\del_-X$ to make the expression transform as $(1,1)$.
The residual transformations maintaining the conformal gauge result in the invariance of these actions
under
\begin{equation}
\label{covtrans}
\delta_\pm~X^\mu = -\epsilon^\pm(\tau^\pm)\del_\pm X^\mu
\end{equation}
$L$ transforms as a $(1,1)$-tensor
and $L^{-1}$
as a $(-1,-1)$ tensor.  
The non-vanishing components of the
Christoffel connection 
are:
\begin{equation}
\label{Gamma1}
{\Gamma^{(1)}}^+_{++} = \del_+ \ln L\quad{\Gamma^{(1)}}^-_{--} = \del_- \ln L\quad D_\pm L =0
\end{equation}
The last of \er{Gamma1} is just the \emph{covariant constancy} of the metric tensor.
We give explicit expressions for some covariant derivatives of interest to this paper:
\begin{eqnarray}
\label{covderivs}
D_\pm~X^\mu &=& \del_\pm~X^\mu\nn\\
D_{++}X^\mu &=& \del_{++}X^\mu - \del_+\ln L\del_+ X^\mu\nn\\
D_{--}X^\mu& =& \del_{--}X^\mu - \del_-\ln L\del_-X^\mu\nn\\
D_{+-} X^\mu &=& D_{-+}X^\mu = \del_{+-}X^\mu
\end{eqnarray}
If $T_{\mu_1\dots\mu_n}$ is a tensor with $m_\pm$ indices of type $\pm$,
\begin{equation}
\label{imprel}
D_{\pm}T_{\mu_1\dots\mu_n} = \del_\pm~T_{\mu_1\dots\mu_n} - m_\pm\del_\pm\ln L  T_{\mu_1\dots\mu_n}
\end{equation}
Another important property is that
$D_{\pm\pm}X\cdot D_\pm X$ are linear combinations of the \emph{gauge fixing conditions}
$g_\pm =\del_\pm X\cdot\del_\pm X$ and their derivatives. 

The manifestly covariant Drummond actions we had analysed in \cite{drumacts} are linear combinations of $S_1^D,S_2^D$
where
\begin{equation}
S_i^D = ~\int \td\tau^+\td\tau^-{\cal M}^D_i  
\end{equation}
where
\begin{eqnarray}
\label{drumlags}
{\cal M}_1^D &=&
{\frac{{\left(D_{++}X \cdot D_{--}X \right)}^{2}}{L^{3}}}\\  
{\cal M}_2^D &=&
{ \frac{\left(D_{++}X \cdot D_{++}X\right) \left( D_{--}X \cdot D_{--}X \right)}{L^{3}}} 
\end{eqnarray}
It was shown there that these actions are scalar densities under \er{urtrans} and they have vanishing \emph{on-shell} $T_{--}$.
The main result of this paper is in extending these results to \emph{all} $S^j_{cov}$.

\subsection{Impossibility of Covariantizing $S_\beta$}
Writing 
$S_\beta = \frac{\beta}{4\pi}\int \td\tau^+\td\tau^- {L}_2$
it is easy to show that
\begin{equation}
\label{psanom}
\delta^{0}_- L_2 = \del_-(\epsilon^- L_2) + \del_-^2\epsilon^- \del_+ \ln L
\end{equation}
If $L_2$ had
transformed as a scalar density, only
the first term would have been there. The presence of the second term means that $S_\beta$ is not manifestly covariant,
though it is \emph{invariant} under \er{urtrans} as the second term is a total divergence. All our efforts at explicitly
covariantizing this action by replacing ordinary derivatives by covariant derivatives only led to trivial results. This, as discussed at length in \cite{covariant}
can be traced to the fact that $S_\beta$ descends from the \emph{Liouville Equation} which is the conformal gauge value of 
the \emph{Polyakov} action($R(\xi)$ is the Ricci scalar)
\begin{equation}
\label{PolAct}
S_{\textup{Polya}} = \int \td^2 \xi \sqrt{g(\xi)} R(\xi) (\frac{1}{\nabla^2} R)(\xi)
\end{equation}
This is the so called WZNW effective action for
conformal anomaly in two dimensions.
In \cite{covariant} we had given a general proof that for WZNW effective actions of \er{PolAct}
the integrand can never be manifestly covariant. The basic reason is that the solution $\phi(\xi)$ to
\begin{equation}
\label{almostscalar}
 \nabla^2 \phi(\xi) = R(\xi) 
\end{equation}
is not a \emph{scalar} even though $R$ is and $\nabla^2$ is the \emph{scalar Laplacian}. In fact $\phi(\xi)=\ln L$.
A related two-dimensional peculiarity is that the equation $\del_\alpha f(\xi) = V_\alpha$ where $V_\alpha$ transforms
as a vector, admits non-scalar solutions for $f(\xi)$.

As happens in the theory of anomalous effective actions, after isolating $S_\beta$ in \er{total}, all other terms will
be \emph{manifestly covariant}. Because of this special nature of $S_\beta$, the analysis of its spectral content has to
use techniques other than what is presented in \ref{sec_gen}. This has been done in \cite{liou-all}.
\section{General Analysis}
\label{sec_gen}
Our analysis begins with a derivation of the full EOM and $T_{--}$ corresponding to the total action $S_{tot}$ of \er{total}.
These are given by
\begin{equation}
\label{defs}
E^\mu = \frac{\delta S}{\delta X^\mu}=0\quad\quad \delta S = \frac{1}{2\pi}\int \td^2~\del_+\epsilon(\tau^-,\tau^+)~T_{--}
\end{equation}
the second part of which is obtained by the usual N\"oether procedure after treating $\epsilon(\tau^-)$ as depending on 
$\tau^+$.
The following relation between the \emph{off-shell} $T_{--}$ and $E^\mu$ is an important one:
\begin{equation}
\label{eomT}
\del_+~T_{--} = -2\pi E\cdot\del_-X
\end{equation}
We now introduce the decomposition 
\begin{equation}
\label{holosplit}
X^\mu = X^\mu_{cl}+F^\mu(\tau^+)+G^\mu(\tau^-) + H^\mu(\tau^+,\tau^-)
\end{equation}
where 
${X^\mu}_{\textup{cl}} = e^\mu_+R\tau^+ + e^\mu_- R \tau^- (e_-^2=e_+^2=0~ \rm{and}~ e_+\cdot e_- = -1/2)$ \cite{ps}
is a classical solution of $S_0$;
$F,G$ are anti-holomorphic and holomorphic functions respectively. $H$ is purely \emph{non-holomorphic} and by construction
it does not have any purely holomorphic and anti-holomorphic parts. 
The form of the full EOM given in \er{defs} 
can be iteratively solved to give $H$ in terms of $F_+,G_-$ and their higher derivatives.
The advantage of introducing the decomposition of \er{holosplit} is that
\emph{off-shell} $T_{--}$ of \er{defs} can be uniquely split as
\begin{equation}
\label{Tsplit}
T_{--} = T_{--}^h + T_{--}^{nh}
\end{equation}
where $T_{--}^{nh}$ is \emph{purely non-holomorphic} in the sense that it has no holomorphic parts.
The \emph{on-shell} $T_{--}$ can be obtained by simply setting 
$T_{--}^{nh} =0$ which follows as a consequence of \er{eomT}. It is only $T^h_{--}$ that is relevant for the spectrum of
the theory.
\subsection{Analysis of $S^j_{cov}$}
It is easy to see that the only way to get a $T^h_{--}$ is for the Lagrangean to be made up of two factors one
of which, called ${\cal L}^h$ has holomorphic terms in it and another, called ${\cal L}^{hvar}$ whose \emph{N\"oether
variation} has holomorphic parts in it. We first enlist possible candidates for ${\cal L}^h$. To this end we note that
all the covariant derivatives $(D_-)^nX^\mu$ are such that they have holomorphic parts. Though it has a +-derivative,
$D_+X = Re_+\ldots$ actually has a holomorphic part. As can be seen from using \er{holosplit} in \er{covderivs} none of
the higher covariant derivatives $(D_+)^n~X$ contains holomorphic pieces. It then follows that no tensor with two or
more +-indices contains holomorphic pieces. Likewise $D_{+-}X$ and all its higher covariant derivatives also do not
contain any holomorphic pieces. In these conclusions we have used the fact that $L~ \rm{and}~ \del_-L$ contain holomorphic
pieces but not $\del_+L$.

To construct ${\cal L}^h$ one forms scalar(in target space) dot products among all the tensors containing holomorphic
pieces; at first sight $D_+X\cdot(D_-)^nX$ appears to be one such. Consider
\begin{equation}
(D_-)^{n-1}L = (D_-)^nX\cdot D_+X+.....
\end{equation}
where the additional terms contain $D_{+-}X$ and its higher derivatives. Since the lhs vanishes due to $D_-L=0$,
and all additional terms are not holomorphic, one concludes that actually $D_+X\cdot(D_-)^nX$ does not contain
holomorphic pieces. Thus ${\cal L}^h$ can contain $(D_-)^mX\cdot(D_-)^nX$ and arbitrary products of such terms.
Of these $D_-X\cdot D_-X$ and $D_-X\cdot D_{--}X$ should not be considered as they are proportional to 
$g_{--}$ and its derivatives, which should vanish in the conformal gauge. This means that every term in ${\cal L}^h$ has \emph{four or more} negative indices.

Constructing ${\cal L}^{hvar}$ involves a little work. To that end the following N\"oether variations are useful:
\begin{eqnarray}
\label{Nvars}
\delta D_+X^\mu &=&\del_+(\delta X^\mu) = \del_+\epsilon^-\del_-X^\mu+\epsilon^-\del_{+-}X^\mu\nn\\
\delta\ln L &=&\del_-\epsilon^-+\epsilon^- \del_-\ln L+\frac{\del_+\epsilon^-}{ L} \del_-X\cdot \del_-X
\end{eqnarray}
From this it follows that among all the higher rank covariant derivatives only $D_{++}X,D_{+-}X$, and their $D_-$
derivatives, can have holomorphic N\"oether variations.
It is also clear that the integrand of $S^j_{cov}$ \emph{can not contain} more than one higher(than
rank one in +) covariant derivative as a factor. The way to construct such integrands is to string together target space
scalar products with equal number of $(+,-)$ indices and divide by sufficient numbers of $L=D_+X\cdot D_-X$(it was for
this reason we did not include $L$ among ${\cal L}^h$). But this would require an element of ${\cal L}^{hvar}$ to have 
at least four \emph{net} +-indices. Every scalar product must necessarily include either a $D_+X$ or $(D_-)^nX$ factor. It
is clearly optimal to choose a $D_+X$ factor. Thus the \emph{single} higher rank
covariant derivative factor must involve $D_{+++}X$ or higher +-derivatives. But none of them can contain a holomorphic
N\"oether variation.

Thus we come to the startling conclusion that for each of the $S^j_{cov}$ there is no holomorphic part to $T_{--}$ to
all orders!
In other words they can not change the spectrum from what the $S_0,S_\beta$ actions give. But we have already shown
in \cite{liou-all} that $S_\beta$ does not correct the spectrum of $S_0$ to all orders. Hence the conclusion of this paper,
namely, \emph{all} conformally invariant effective string theories are \emph{isospectral} to the Nambu-Goto theory, $S_0$.
\section{Discussion and Conclusions}
In this paper we have shown that the entire class of conformally invariant Polchinski-Strominger effective string theories 
has the same spectrum as Nambu-Goto theory to all orders in $R^{-1}$, where $2\pi R$ is the length of the closed string.
The proof consisted of two steps the first in demonstrating that $S_\beta$, though contributing non-trivially to the on-shell
$T_{--}$ nevertheless does not correct the Nambu-Goto spectrum \cite{liou-all}; next it is shown that no manifestly covariant
action even contributes to the on-shell $T_{--}$. An immediate comparison can be made with the results of Aharony and Karzbrun
\cite{aharony} who, following the L\"uscher-Weisz approach, showed similar results to order $R^{-5}$ in $D=3$ but claimed
that for $D\ge 4$ there could be corrections. This discrepancy between our results based on the PS-formalism,
and their results based on the LW-formalism needs to be understood. If both our calculations are correct, it may point to the
possibility that \emph{conformal invariance} in addition to ensuring only $D-2$ transverse degrees of freedom, may also be
restricting the interactions. It is very important to reexamine whether such features are justifiable from the point of view
of the relevant microscopic theory, for example QCD. In other words, the \emph{symmetry content} of effective string theories needs a fresh look.
But such a fresh analysis has to retain the agreement between numerical studies and Nambu-Goto spectrum to order $R^{-3}$. First principle derivation of the effective actions as done by Akhmedov et al could throw
valuable light on these issues \cite{emil}. Numerical data \cite{ournum} clearly shows a deviation from Nambu-Goto theory 
at intermediate distances. For $D=4 SU(3)$ QCD this was around 0.75fm.  This is also the situation with numerical simulation of percolation models \cite{perco}. One possibility
of reconciling this is if at these intermediate scales the string-like object has not formed at all. If not, one will have to
conclude that conformally invariant effective string theories do not provide a good description. Our analysis does not seem
to provide any room for \emph{extrinsic curvature string} effects. Once again it should be stressed that the Polyakov action 
\cite{extrinsic} for such strings does not have the conformal invariance used here. It is also very important to find out what
additional physics is coded by the large class of highly non-trivial actions considered here. Studies of scattering amplitudes
and Partition functions ( as studied in \cite{aharony}) may be needed for that. It is also desirable to get a deeper physical 
understanding of our results. It is clear that we are still a long way from understanding QCD-Strings.

\section*{Acknowledgements}
The author gratefully acknowledges the DAE Raja Ramanna Fellowship Scheme under which this work was carried
out, and CHEP, IISc, Bangalore, for facilities. He thanks H. Kawai, T. Yukawa , A. 
Sen, V. Vyas and G. Bali for many valuable discussions. 
Special thanks are to P. Matlock for an extended collaboration 
, and to Y. Bharadwaj for verifying many results and for finding better
ways of deriving them.


\end{document}